# Probing the Turbulence Dissipation Range and Magnetic Field Strengths in Molecular Clouds II: Directly Probing the Ion-Neutral Decoupling Scale

Kwok Sun Tang,[1] Hua-Bai Li,[1] and Wing-Kit Lee[2]

[1]*Department of Physics, The Chinese University of Hong Kong, NT. Hong Kong SAR*
[2]*Institute of Astronomy and Astrophysics, Academia Sinica, P.O. Box 23-141, Taipei 10617, Taiwan*

## ABSTRACT

The linewidth of ions has been observed to be systematically narrower than that of the coexisting neutrals in molecular clouds (Houde et al. (2000)) and been interpreted as the signature of the decoupling of the neutral turbulence from magnetic fields in partially ionized medium (Li & Houde (2008); Paper I, hereafter). As a sequel of Paper I, here we present further observational evidence that lend support to these earlier proposals with the velocity coordinate spectrum analysis (Lazarian & Pogosyan (2006)). We recover the turbulent energy spectra of $HCN$ and $HCO^+(4-3)$ in a starless molecular filament in NGC 6334 where magnetic fields play a dynamically important role (Li et al. (2015)). Our analysis showed that the neutral spectrum is consistent with Kolmogorov-type ($k^{-5/3}$, where $k$ is the wave number), while that of the ions is the same on the large scale but steeper ($\sim k^{-2}$) for scales smaller than $0.404pc$. We carefully ruled out the possibilities that the spectrum difference can stem from the differences of ion and neutral optical depth and hyper-fine structure.

Corresponding author: Hua-Bai Li, Kwok Sun Tang
hbli@phy.cuhk.edu.hk, kwoksun2@illinois.edu

1. INTRODUCTION

Magnetic fields are ubiquitous in the Milky Way and are believed to play various important roles in converting galactic gas into stars (For a recent review see Li et al. (2014)). In most analyses and numerical simulations, gas and B-fields are treated as perfectly coupled. This is a fairly good approximation when large scales are considered. However, when one approaches smaller scale, the decoupling between gas flows and B-fields, also known as Ambipolar Diffusion (AD), becomes inevitable. AD is the physical process that describes the relative motion of neutrals and charged particles attached to magnetic field lines in partially ionized media. In the context of star formation, it manifests itself in three different flavors, depending on how the relative motion between the two species is being driven.Gravitationally-driven AD, first identified by Mestel & Spitzer (1956) and further developed by Mouschovias (1979); Nakano et al. (2002), corresponds to the process of how neutrals drift through the charged particles due to its self-gravity, and how it consequently leads to the increase in the mass-to-magnetic-flux ratio of the molecular clouds. Magnetically-driven AD (Mouschovias (1987); Mouschovias et al. (2011)) is concerned with the process of how magneto-hydrodynamic (MHD) waves are damped due to the imperfect coupling between neutrals and charged particles in non-turbulent media. It has been proposed to facilitate the removal of turbulent support against gravity and possibly initiate the fragmentation of molecular cloud cores (Mouschovias (1987, 1991); Mouschovias et al. (2011)). Turbulence-driven AD (Paper I,Tilley & Balsara (2010); Burkhart et al. (2015)), on the other hand, is concerned with the damping of perturbations in B-fields in turbulent, partially-ionized media. In contrast to the magnetically-driven AD, turbulence-driven AD predicts that the turbulent energy/ motions in the neutrals would not be damped as soon as the scale at which neutrals decouple from the B-fields is reached. Instead, small eddies in the neutrals would be fed by its own turbulent energy cascades, while the turbulent motions of the ions, which is tied to the B-fields, would be dissipated (Paper I).

Despite its importance in various aspect of star formation, to date, there is no direct observational evidence for any kinds of AD. Observations of AD are either indirect (Paper I, Hezareh et al. (2010, 2014)) and/or highly controversial (Crutcher et al. (2010); Mouschovias & Tassis (2010); Li et al. (2014); Tritsis et al. (2015), Jiang et al. 2018).

In Paper I, we proposed that the linewidth difference between coexistent ions and neutrals is the manifestation of AD in turbulent medium, and can be observed even when the ion-neutral decoupling scale $l_{AD}$ is unresolved by the telescope beam. The rationale behind it is that at scales smaller than $l_{AD}$, only ions are coupled to the magnetic fields, therefore their turbulent motions are inevitably damped in the presence of turbulence-driven AD. The low ionization level of molecular clouds limit the frictional force that acts on the neutrals by ion-neutral collisions. Small-scale motions of neutrals would still be sustained by its own turbulent energy cascade until hydrodynamic viscosity sets in. This claim is further developed and verified by later work like Li et al. (2010); Hezareh et al. (2010, 2014). Although this theory readily explained the narrower linewidth of $HCO^+$ than that of consistent $HCN$, as first pointed out in Paper I, the turbulent linewidth-size relation reported ($\Delta v(k) \propto l^{0.15}$) are consistently shallower than what is expected theoretically for either the Kolmorgorov-type incompressible turbulence $l^{1/3}$ or the observationally favored compressible turbulence $l^{1/2}$ (Heyer & Brunt (2004); McKee & Ostriker (2007)) where $l$ is the size of the turbulent eddies considered. Such discrepancies between theory and observations may propagate and eventually affect the determination of the $l_{AD}$. On one hand, turbulence-driven AD, as we will show in the next section, predicts an $l_{AD}$ at sub-pc scale, the observations seem to support an $l_{AD}$ of the order $\sim 1-10 mpc$ (Paper I, Hezareh et al. (2010, 2014)). On the other hand, observational biases such as the optical depths, relative abundances of the tracers $HCN$ and $HCO^+$ are often being criticized as the culprit for the observed linewidth difference between the ions and neutrals.

In this work, we revisit turbulence-driven AD with the velocity coordinate spectrum (VCS) analysis to our observed $HCN$ and $HCO^+$ spectral data cubes. This technique enables the modeling of the turbulent energy spectra of ions and neutrals, as it can disentangle the underlying turbulence statistics from the contributions of density structures to the spectral lines. This method has been successful in retrieving statistical description of turbulence in simulation data cubes (Chepurnov et al. (2010); Padoan et al. (2009)) and has been applied to various observational data (Chepurnov et al. (2015); Padoan et al. (2009)). We present our observation and analysis towards our carefully chosen diffuse molecular filaments in NGC 6334 where magnetic fields are shown to be dynamically important. We will justify that neither the optical depths nor other intrinsic properties of the tracers are responsible for our results.

We begin with an introduction of turbulence-driven AD and the physics of neutral-ion decoupling, and show that it is a relevant process based on widely observed scaling relation on the scale of molecular cloud cores in Section 2. Data acquisition and the rationale behind target selection is presented in Section 3. It is followed by an overview on the VCS analysis (Lazarian & Pogosyan (2006)), and the application on NGC 6334 in Section 4 and 5 respectively. We carefully rule out the factors that might influence our analysis in Section 6. In Section 7, we repeat the same analysis as in Paper I and discuss the origin of the consistently shallower turbulent spectrum presented by these previous works. Finally, the implications of our observation results are elaborated in Section 8, followed by a conclusion in the final Section 9.

## 2. THE PHYSICS OF TURBULENCE-DRIVEN AMBIPOLAR DIFFUSION

To understand turbulence-driven AD, we have to first make sense of how turbulent flows, carried mostly by neutrals, are coupled to the B-fields through their collision with ions. The Lorentz force can only be felt by charged species. For neutrals to stay magnetized, collisions between neutrals and ions must be frequent enough for neutrals to capture the motion of ions coupled with the B-fields. The Magnetic Reynold number $R_M$ is a dimensionless parameter that characterizes how well these two species are coupled for turbulent motion of different scales (Paper I, Balsara (1996); Oishi & Mac Low (2006); Tilley & Balsara (2010); Meyer et al. (2014)). At scales where $R_M \gg 1$, neutrals and ions are well-coupled. When $R_M$ is small, neutrals are no longer frozen into the B-fields and decouple from the charged species which remain attach to the B-field lines. As a result of the decoupling, ions and neutrals should demonstrate distinct turbulent energy spectra for scales smaller than $l_{AD}$ as shown in Figure 1A. The imperfect coupling between the two species induces friction forces on each other's flow and leads to the damping of motions for scales below $l_{AD}$. Due to the low ionization level in molecular clouds, charged particles are outnumbered by neutral particles by almost ten million times. This means that a charged particle encountering neutral particles is far more likely to occur than a neutral particle bumping into charged particles. Therefore, small-scale motion of ions are expected to be heavily damped in this regime, and this will then prohibit the cascade of turbulent energy further into even smaller lengthscale. The frictional force between ions and neutrals is expected to play a role in damping the turbulent energy of neutrals as well, however a minor one due to the low ionization level (Paper I). Since neutrals decouple from the magnetic fields and ions, small scale turbulent motion in neutrals are still fed on its own turbulent energy cascade and should survive beyond the decoupling scale. The energy cutoff in neutrals would still predominantly be set by hydrodynamic viscosity for scales smaller than the mean free path of the fluid (Paper I). A rough estimate of the decoupling scale $l_{AD}$ can be made by setting $R_M = 1$:

$$R_M = \frac{l_{AD} V_n}{\beta} = 4\pi m_i m_n \gamma_d \frac{\chi_i n_n^2 l_{AD} V_n}{B^2} = 1,$$

$$l_{AD} = \frac{B^2}{4\pi m_i m_n \gamma_d \chi_i n_n^2 V_n(l_{AD})}$$

(1)

where $V_n$ is the characteristic velocity at the scale of $l_{AD}$, $\beta$ is the effective magnetic diffusivity and is given by $\beta = B^2/4\pi m_i m_n \gamma_d \chi_i n_n^2$ (Paper I, Balsara (1996)) when AD dominates over other non-ideal processes like Hall effect and Ohmic dissipation. Here $n_n$ is the number density of neutrals and $\gamma_d$ represent the drag coefficient arising from the exchange of momentum between ions and neutrals. For typical conditions in molecular clouds, mean neutral $m_n$ and ion mass $m_i$ are 2.3 and 29 times that of the hydrogen atom, and the drag coefficient $\gamma_d$ is estimated to be $1.5 \times 10^{-9} cm^3 s^{-1}$) (Paper I, Mouschovias et al. (2011)). $\chi_i$ is the ionization fraction of the molecular clouds. Assuming the linewidth-size relation $\sigma \approx 0.9 kms^{-1}(\frac{l}{1pc})^{0.5}$ (Heyer & Brunt (2004)) and B-field-density scaling relations $B \approx 10\mu G(n/(300cm^{-3}))^{0.65}$ (Crutcher et al. (2010)), or $B = 0.19mG(n/(10^3 cm^{-3}))^{0.41}$ (Li et al. (2015)), one would expect that the decoupling of neutrals from ions should be at work already at sub-pc scale $l_{AD} \approx 0.13 - 0.41 pc$ with typical cloud cores density $n \sim 10^4 cm^{-3}$ where ionization fraction $\chi_i$ is approximately $10^{-7}$ for an assumed cosmic ray ionization rate of $\approx 10^{-16} s^{-1}$(Draine (2011)).

The decoupling scale predicted by magnetically-driven AD (Mouschovias (1987); Mouschovias et al. (2011)) is comparable to the $l_{AD}$ presented in this work. However, such models predict that motion of neutrals beyond decoupling scale would also be damped rapidly, and therefore require energy sources other than MHD waves to explain the observed non-thermal neutral linewidth (Paper I, Hezareh et al. (2010, 2014)). Furthermore, two-fluid turbulent MHD simulations (Tilley & Balsara (2010); Burkhart et al. (2015)) have also shown that the decoupling is characterized by $l_{AD}$ and they lend further support to the picture that neutral turbulence are maintained beyond $l_{AD}$. We caution our readers that turbulent-driven AD presented here should not be confused with turbulence-accelerated AD (Zweibel (2015); Fatuzzo & Adams (2002)) which describes how turbulence help redistribute the magnetic flux in star-forming clouds, and reduce the gravitationally-driven AD timescale.

## 3. TARGET SELECTION AND DATA ACQUISITION

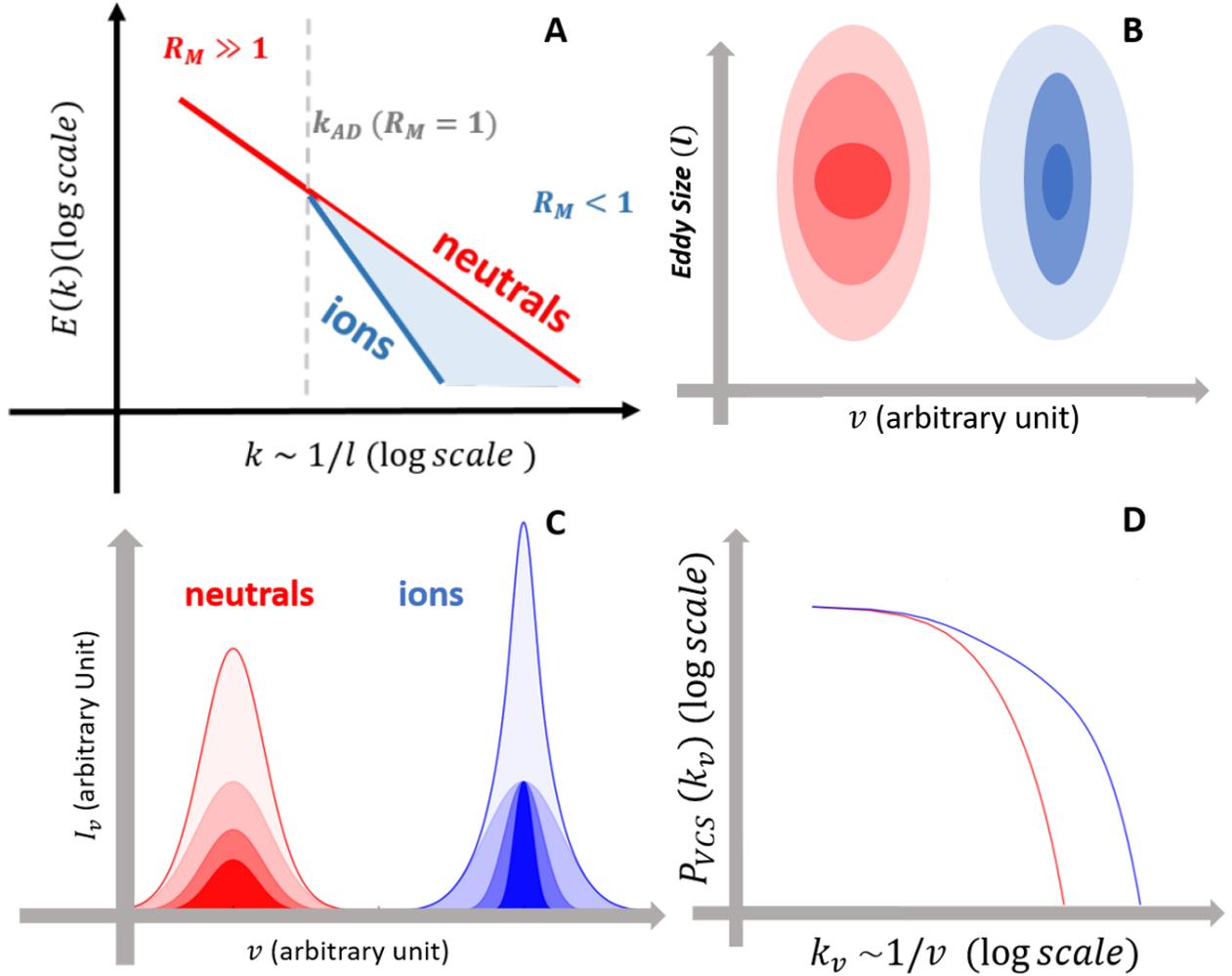

**Figure 1.** Panel A: A Log-Log plot of the turbulent velocity spectrum (energy density per unit wavenumber and mass) versus wavenumber in arbitrary units for demonstrative purposes. The velocity dispersion/linewidth of the tracers measured at scale $L$ is proportional to the integration of the turbulent energies contributed by all eddies of size smaller than $L$. The turbulent energy in ions is dissipated steeply for scales smaller than $l_{AD}$ because of turbulence-driven AD. Panel B: An illustration of mapping turbulent eddies from real space to spectral lines (velocity space) inspired by (Lazarian (2009)). In real space, there are 3 eddies of same density, but different velocity dispersion. The larger the velocity dispersion of eddies, the larger the extent of it in the velocity space. Here the largest eddy of these two species share the same velocity dispersion. Due to the difference in the turbulent energy spectra (Panel A), the energy carried by the small ions eddies decreases faster with scale. The spread of these blue small eddies (ions) in the velocity axis is smaller than their counterpart in red (neutrals). Panel C: We model the spread of the eddies in the spectral space as Gaussian profile with the width being their respective velocity dispersions. The area under these Gaussians is scaled linearly with the size of the eddy l while the velocity dispersions of the red curve and blue curve are scaled with $\sigma_v \propto l^{1/3}$ and $\sigma_v \propto l$ to imitate the difference in turbulence energy spectrum. The eventual spectral line (solid line) is given by the collective contributions of these profiles from individual eddies. Panel D: The mocked VCSs of the spectral lines modeled in Panel C. For neutrals, the intensity fluctuations are more widely spread in the velocity space, while for ions, the intensity fluctuations are more localized in the narrow band of velocity channel because turbulence-driven AD would tend to dissipate energy in small eddies. For the species with less energy in small eddies, the VCS of its spectral lines should therefore contain more power in large $k_v$ space. By studying the behavior in the Fourier space, VCS recovers the statistical description of the turbulence spectrum.

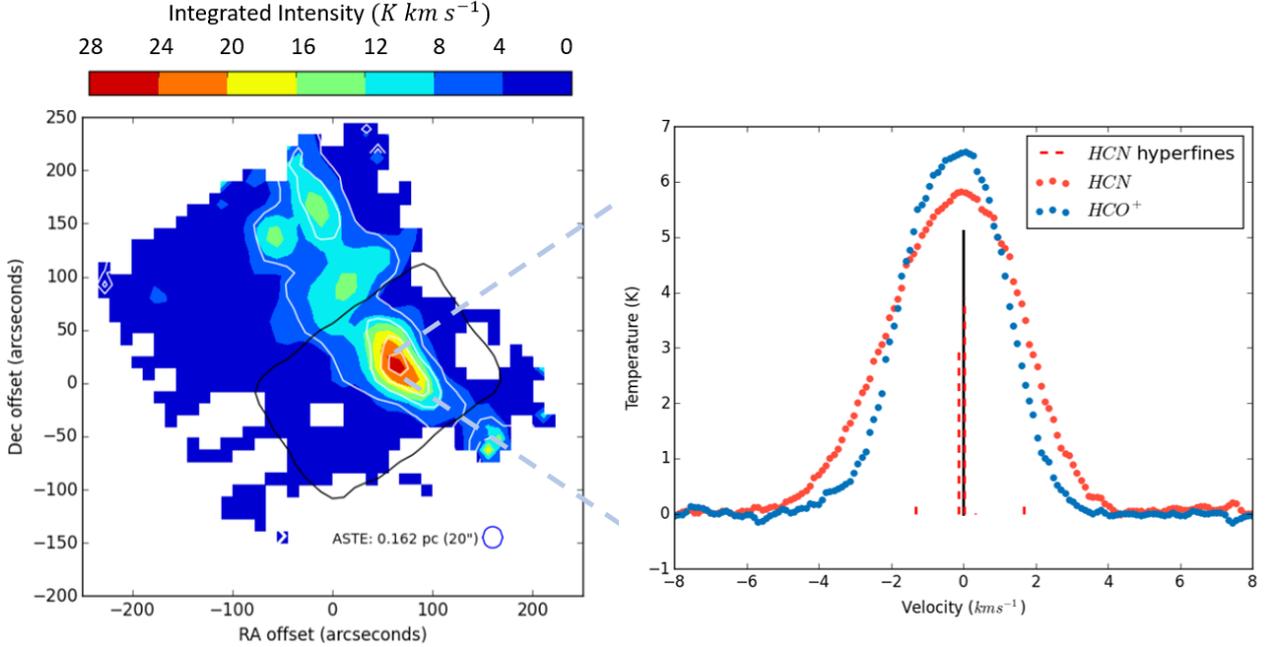

**Figure 2.** Left: Integrated Intensity maps of $HCN$ (white contours) and $HCO^+$ (colored plot) of this filament in NGC6334. The reference coordinate of this map is $17h20m36.8s, -35°51'26''$. The highest contour levels correspond to an integrated intensity of $20 Kkms^{-1}$; each following level is lowered by $5 Kkms^{-1}$ accordingly. The colorbar is in the scale of $Kkms^{-1}$. The region adopted for VCS analysis is marked with a black line. Right: The spectral line profiles of both $HCN$ and $HCO^+$ from the densest region show no signs of self-absorption. And the relative intensity of $HCN$ hyperfines under LTE (26) are plotted in dashed lines. The central and brightest component is shown in the grey solid line. The two brightest hyperfine satellites are only barely resolvable in our observations.

We chose a molecular filament in NGC 6334 where B-fields have been shown to lie close to the plane of sky and play a dynamically important role in regulating the filaments fragmentation (Li et al. (2015)). Complementing this with the fact that the filament is situated far from regions with stellar feedbacks (NGC 6334 I, IN, V), it is an ideal venue for us to scrutinize this phenomenon in closer detail. Additionally, this relatively diffuse filament ($n(H) \sim 10^4 cm^{-3}$) ensures that our tracers ($HCN(4-3)$, $HCO^+(4-3)$) are optically thin (Section 6), and the observed line profiles are not broadened by the optical opacity (Right panel of Figure 2).

$HCN(4-3)$ and $HCO^+$ are chosen as they are observed to be well-correlated spatially in previous observations towards star forming regions of different environments (e.g. Houde et al. (2000); Lo et al. (2009)). The observations of this filament were made with the Atacama Submillimeter Telescope Experiment (ASTE). With ASTE's 20" resolution, the physical size of the achievable resolution is about $0.16pc$ assuming NGC 6334 is $\sim 1.7kpc$ away (Russeil et al. (2012)). The left panel of Figure 2 shows the integrated intensity map of both $HCN$ and $HCO^+$; the contour ($HCN$) and the color plot ($HCO^+$) are highly correlated spatially. This indicates that we are comparing neutrals and ions that trace the similar volume of the molecular clouds. For our carefully chosen filament, neither the enhancement due to outflows nor the chemical evolution in proto-stellar cores would give conceivable change on the abundances of this neutral-ion pair (See Section 6).

## 4. METHOD: VELOCITY COORDINATE SPECTRUM

The Velocity Coordinate Spectrum (VCS) is defined as the power spectrum of the intensity fluctuations in the velocity coordinate along fixed line of sight:

$$P_{vcs}(k_v) = |\int_{-\infty}^{\infty} I(v) e^{ik_v v} dv|^2 \qquad (2)$$

| $l$ | $HCN$ | $HCO^+$ |
|---|---|---|
| $l_{AD} < l < L$ <br> $L = 1.26 \pm 0.11 pc$ | $E(k) = (20.0 \pm 4.3)k^{-1.66 \pm 0.09} km^2 s^{-2} pc^{1.66}$ | |
| $l < l_{AD}$ <br> $l_{AD} = 0.404 \pm 0.132 pc$ | same as above | $(49.7 \pm 13.6)k^{-2.01 \pm 0.05}$ <br> $km^2 s^{-2} pc^{2.01}$ |

**Table 1.** Best-fit turbulent energy spectrum per unit mass returned from fitting VCS to the observed power spectrum of $HCN$ and $HCO^+$.

where $I(v)$ is the spectral line intensity at velocity $v$ and $k_v$ is the wavenumber in the velocity coordinate (i.e., $k_v = 2\pi/v$). The theoretical prediction of such a power spectrum is backed by the assumption of a power law like energy spectrum for both turbulent velocity distribution, i.e. $E_v(k) \propto k^{-\kappa}$ and the density distribution $E_\rho(k) \propto k^{-\gamma}$ where $k$ corresponds to the wave number in spatial coordinate (i.e. $k = 2\pi/l$), $\kappa$ refers to the turbulence spectral index. For Kolmogorov-type turbulence, the spectrum is expected to show a scaling of $\kappa = 5/3$. The full functional form of VCS depends further on temperature $T$, VCS spectrum amplitude $P_0$, noise level $N_0$, and the size of the cloud $L$.

$$P_{vcs}(k_v) = P_{vcs}(E_\rho(k), E_v(k), T, P_0, N_0, L) \quad (3)$$

The non-trivial nature of the turbulent energy spectrum and its contribution to the intensity fluctuations along the velocity coordinate can be visualized using the schematic in Figure 1B and 1C inspired by Lazarian (2009). Assuming three eddies of similar density but different velocity dispersion. The largest eddy would span the largest extent in the velocity axis, and therefore its contribution to the intensity fluctuations would be relatively evenly spread on the resultant spectral lines. On the other hand, the least energetic eddy would be localized in the velocity space and dominates the intensity fluctuations in the spectral lines. As a result of turbulence-driven AD, the small eddies of the ions should carry less turbulent energy compared to that of the neutrals, and contribute its intensity across a narrower velocity range (Figure 1B). The ion spectral line is therefore expected to be more sharply peaked (Figure 1C). In the Fourier space, this translates into a shallower power spectrum (VCS) as the intensity fluctuations are predominantly concentrated in the larger $k_v$ space (Figure 1D).

The power spectrum of $HCN$ and $HCO^+$ (taken along the velocity axis) are first averaged over all the line of sights to lower the noise. We focus our analyses towards the core region of this filament (the region enclosed by the black contour in Figure 2), as other regions of this filament have multiple density structures along their velocity axis. In light of the description of the schematic Figure 1 and the simulation results from Meyer et al. (2014); Burkhart et al. (2015); Li et al. (2012), the turbulent spectrum of $HCO^+$ cannot be simply described by a single spectral index turbulent spectrum. Therefore, in our modelling of VCS, we have included one more spectral index $\kappa_{AD}$ to describe the damping effect induced by magnetically-driven AD for scales smaller than $l_{AD}$ (i.e. $\kappa_{AD} > \kappa$). We should also note that in two-fluid turbulent MHD simulations, the transition of ions power spectrum from larger scale to the decoupling scale is usually smooth compared to our broken power-law model (Meyer et al. (2014); Burkhart et al. (2015); Li et al. (2012)). Albeit simplified, our model captures the essential physics of the decoupling: the transition scale $l_{AD}$ and the steep decay of turbulent energy in ions, while at the same time streamline our fitting procedures. The detailed fitting procedures are outlined in the Appendix A where we also show the temperature dependent term and density distribution term play a minor role in modifying the behavior of $P_{vcs}(k_v)$.

## 5. RESULTS

The modelling of VCS begins by first fitting the power spectrum of $HCN$, as it has less input parameters compared to $HCO^+$ and returns the information of the turbulent spectrum regarding larger scales (e.g. $\kappa$, $L$). These parameters are used in the further fitting of the spectrum of $HCO^+$ (e.g. $l_{AD}, \kappa_{AD}$).The resultant best fit parameters and their respective standard deviation returned from the Scipy Optimize module and the error propagation are listed in Table 5.1.

### 5.1. *Turbulent Spectra of ions and neutrals*

The best fit curve is shown together with the VCS of the two species in Figure 3. The best fit parameters in Table 5.1 indicates that $HCO^+$ traces a steeper slope ( $k^{-2.04}$) than that of $HCN$ ( $k^{-1.66}$ ) for $k > k_{AD}$. This corresponds to a steeper decay of turbulent energy of the charged species and agrees with the theoretical prediction. The spectral index obtained for the neutral species is in agreement with ideal MHD simulations ( $k^{-5/3}$ (Cho & Lazarian (2003))) and non-ideal simulations in the limit where $R_{AD}$ is large ($k^{-5/3}$ (Burkhart et al. (2015)), $k^{-3/2}$ (Li et al. (2012)),). Beyond the $l_{AD}$, simulations predicted a steeper decay ($k^{-4}$ (Burkhart et al. (2015))) than what we observed. However, attention must be paid carefully to interpret simulation results as numerical viscosity would also damp the turbulent velocity structures at smaller scales. Higher resolution simulations would be needed to confirm the results in the future.

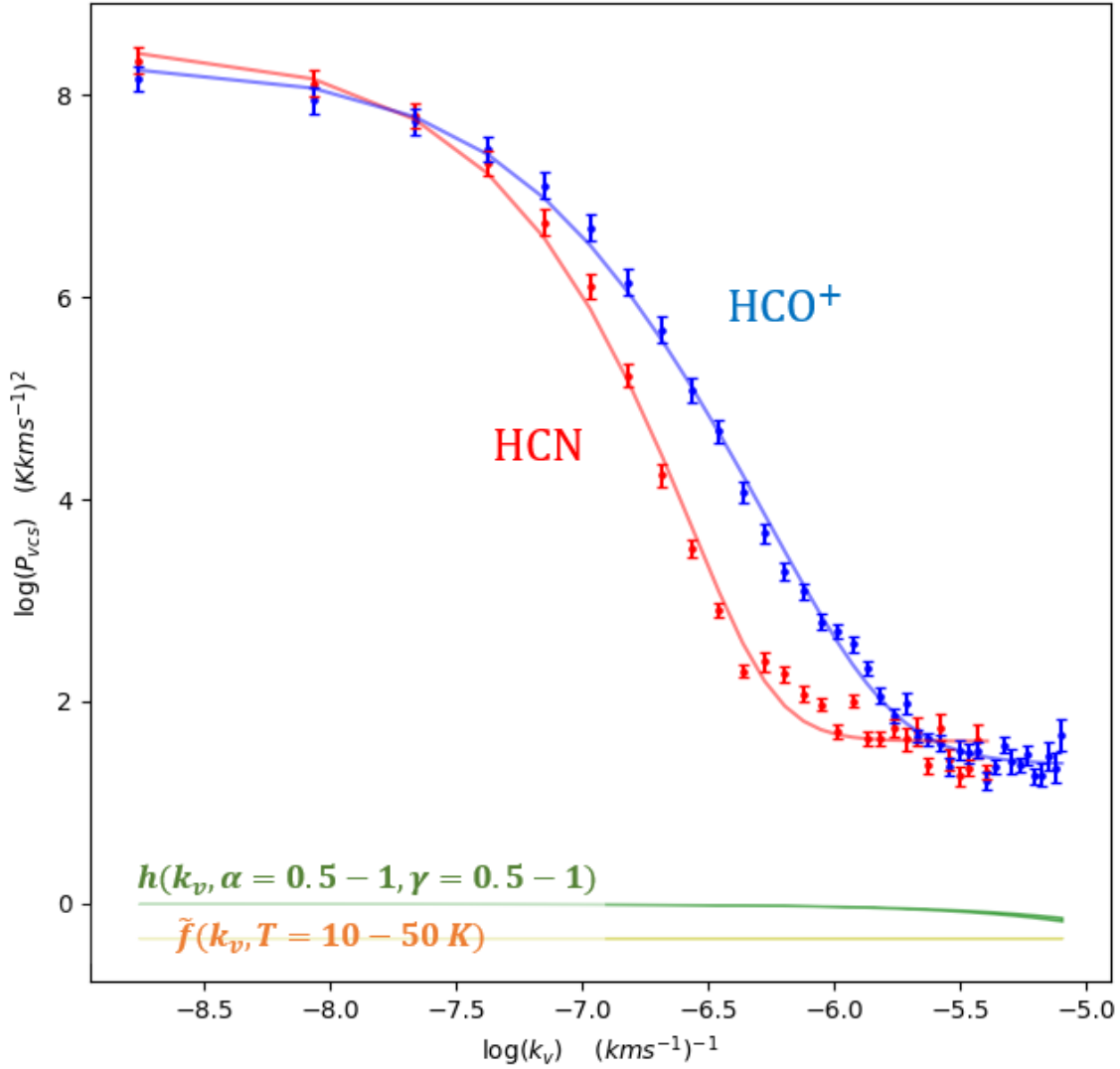

**Figure 3.** The power spectrum obtained for $HCN$ and $HCO^+$ are shown in log-scale. Error bars correspond to one standard deviation in the scatter plot for $HCN$ and $HCO^+$ (red and blue, respectively). The best fits returned from the fitting of VCS with these data are overlaid onto the set of data as solid lines. The effects of both hyperfine structures and the gas temperature enter as extra factors in addition to the VCS, i.e. $P(k_v) = h(k_v, \alpha, \gamma) f(k_v, T) P_{vcs-fit}(k_v)$ (Appendix). Since Figure 3 is presented in log-log space, these extra factors would simply upshift/ downshift our VCS best-fit. The green line demonstrates how the inclusion of the two brightest and most probable hyperfine structures $h(k_v, \alpha, \gamma)$ would affect the behavior of VCS for $\alpha$ and $\gamma$ that ranges from 0.5 to 1, where $\alpha$ and $\gamma$ are the exaggerated relative intensity of these hyperfine satellites to the main line. The yellow line demonstrates the effect of the temperature dependent term $f(k_v, T)$ for temperature ranging from $T = 10 - 50K$. It is nevertheless clear that both temperature and hyperfine satellite would contribute only when large $k_v$ is considered.

### 5.2. Estimating the B-field

Following the recipe of Paper I, the plane of sky B-field strength can also be estimated with the decoupling scale $l_{AD}$ and the turbulent energy spectrum by again setting the magnetic Reynold number to one:

$$B = 0.316 mG \left(\frac{l_{AD}}{0.5pc}\right)^{\frac{1}{2}} \left(\frac{V_n}{1 km s^{-1}}\right)^{\frac{1}{2}} \left(\frac{n_n}{10^4 cm^{-3}}\right) \left(\frac{\chi_i}{10^{-7}}\right)^{\frac{1}{2}} \quad (4)$$

where $V_n$ is the velocity dispersion that corresponds to eddies at the decoupling scale $l_{AD}$. $V_n$ can be estimated by multiplying the energy spectrum with the spread of turbulent eddies in the spectral space $\Delta k$, i.e. $V_n \approx \sqrt{E(k)\Delta k}$ ). Assuming a turbulent eddy of size $l$ is normally distributed on the spatial axis with a full-width-half-maximum of $L$, the corresponding spread in the spectral space would be $\Delta k = \sqrt{8ln2/l}$ (Paper I). This gives $V(l_{AD}) \sim 1.11 \pm 0.27 km s^{-1}$. A recent analysis on filaments in NGC 6334 was performed with a combination of multi-wavelength dust continuum measurements (André et al. (2016)). A column density of $0.7 - 1 \times 10^{23} cm^{-2}$ is estimated for this filament. This gives an average density $n(H_2) \approx (2.19 \pm 0.32) \times 10^4 cm^{-3}$ with our estimated line of sight scale $L \approx 1.26 pc$. For $n_n \approx 10^4 cm^{-3}$, the ionization fraction is approximately $10^{-7}$ for an assumed cosmic ray ionization rate of $\approx 10^{-16} s^{-1}$ (Draine (2011)). Using these values, we arrive at an estimated B-field strength $0.665 \pm 0.185 mG$. We compared this estimate with the $B - n$ scaling relation $B = 0.19 mG(n/(10^3 cm^{-3}))^{0.41}$ from the same region (Li et al. (2015)). For a number density of $n(H) \sim 2.19 \times 10^4 cm^{-3}$, it predicts a B-field strength of $0.673 mG$ which is close to our independent estimate.

## 6. POSSIBLE EFFECTS OF THE TRACERS ON OUR STUDY

### 6.1. Relative Abundances between $HCN$ and $HCO^+$

The chemical evolution of ions and neutrals in pre-stellar cores was studied (Tassis et al. (2012)) and their simulations demonstrated the variation of abundance ratio between $HCN/HCO^+$ spans several orders of magnitude in almost all the models with magnetic fields. However, the effects of turbulence are not included. In the presence of turbulent mixing, the relative abundance should vary to a lesser degree. Worthy to note is that $HCO^+$ is shown to be more abundant at larger spatial scale which traces regions with higher level of turbulence. In principle, it should give rise to a larger linewidth than the neutral species. It is contrary to observations reported ( Paper I,Hezareh et al. (2010, 2014)) as well as the map we presented. In particular, at scales comparable to the size of the cores ($\approx 0.1 pc$), the radial dependence of $HCN/HCO^+$ abundance ratio flattens out (Tassis et al. (2012)). Our observation is incapable of resolving the core scale with a beam size of $\sim 0.16 pc$ and are thus probing scales where the abundance ratio of $HCN$ and $HCO^+$ is roughly uniform. On the other hand, the abundances of both $HCN/HCO^+$ are observed to be boosted in the presence of outflows. $HCN$ traces the most energetic outflows while $HCO^+$ generally traces regions close to the base of the outflows (Walker-Smith et al. (2014)). The former is reported to be more enhanced by a factor of $\sim 10-100$ times in the outflow linewings (Tafalla et al. (2010)). However, the filament we presented in our study is far away from the active star-forming region NGC 6334(I), (IN) and (V) with prominent stellar feedback. No detectable outflow however has been reported in our region of interest.

### 6.2. Hyperfine Structures of HCN

The quadrupole moment induced by the nitrogen atom of HCN is responsible for its widely spaced hyperfine structures. For transitions from rotational level $J = 4 - 3$, there are 5 transitions based on the measurement of (Ahrens et al. (2002)). Assuming local thermal equilibrium condition, the two brightest satellite lines are deviated from the main transition by $0.12 km s^{-1}$ and $0.038 km s^{-1}$ respectively. This offset cannot account for our observed $\sigma$ difference between the two species ($\sim 0.4 km s^{-1}$).

The effect of hyperfine structures on the application of VCS is addressed here. Assume the main line has the form of $I(v)$, the spectral line with the inclusion of the 2 most probable satellite lines can be mimicked by

$$I_{hyper}(v) = \alpha I(v - a) + I(v) + \gamma I(v + b) \quad (5)$$

where $\alpha$ and $\gamma$ are the relative peak intensity of the two satellite lines; $a$ and $b$ are the shift of the hyperfine satellites on the frequency/velocity axis with respect to the main lines. The resultant power spectrum of the spectral lines in the presence of hyperfines would be in the form of:

$$\begin{aligned}P_{hyper}(k_v) &= |\int_{-\infty}^{\infty} I_{hyper}(v)e^{ik_v v} dv|^2 \\ &= [1 + \alpha^2 + \gamma^2 + 2\alpha cos(k_v a) + 2\gamma cos(k_v b) + 2\alpha\gamma cos(k_v(a+b))]P_{vcs}(k_v) \\ &= h(k_v, \alpha, \gamma)P_{vcs}(k_v)\end{aligned} \quad (6)$$

The resultant power spectrum has an additional factor which depends on $k_v$. The effect of including this extra factor $h(k_v, \alpha, \gamma)$ on the VCS is demonstrated in Figure 3. Since Figure 3 is presented in log-log space, the inclusion of $h(k_v, \alpha, \gamma)$ would simply upshift or downshift $P_{VCS}(k_v)$ and will not change the shape of the VCS fitting. The hyperfine factor $h(k_v, \alpha, \gamma)$ stays rather uniform for the range of interested $k_v$ and has minute effects on the modelling of VCS. Similarly, the temperature dependent term can be factored out of the formula for VCS and has insignificant variation over the range of $k_v$ concerned.

### 6.3. Optical Depths of the tracers

We estimate the optical depth of the two tracers $HCN$ and $HCO^+$ with online version of RADEX van der Tak et al. (2007). $HCO^+(3-2)$ data of NGC 6334 is best fitted to the model of filament with fractional abundance of $HCO^+$ $X(HCO^+) = 2 \times 10^{-9}$ ( Zernickel et al. (2013)). Assuming a column density of $N(H) \approx 8.5 \times 10^{22} cm^{-2}$ as above, this gives a column density of $N(HCO^+) \approx 1.7 \times 10^{14} cm^{-2}$. For $(4-3)$ transition, $HCO^+$ is marginally optically thin with an optical depth of 0.86. Although there are no existing measurements of $HCN$ abundances on this particular filament, we can estimate the $N(HCN)$ with the abundance ratio between $HCN$ and $HCO^+$ found in general star-forming regions ( $N(HCN)/N(HCO^+) \approx 2$) (Godard et al. (2010)). This gives an HCN optical depth of 0.127 which again is optical thin. A more detail discussion of the effect of optical depth on VCS is given in (Lazarian & Pogosyan (2006)). They suggested that absorption will tend to diminish the correlation between intensity of emissions at widely separated velocities and line-of-sight scales. They also provide the velocity range v applicable for the application of VCS which is $v < V_{abs} \approx V(L)/\tau$ (we rearranged the equation 72 in (Lazarian & Pogosyan (2008)); $\tau$ is the optical depth of the tracer; $V_{abs}$ is the upper velocity limit set by the effect of absorption). A more sophisticated treatment for self-absorbed lines is given (Lazarian & Pogosyan (2008)). In our particular case, $HCN$ will be optically thinner than $HCO^+$. This means HCN is better at tracing the entire range of turbulent spectrum and will therefore put a good constraint on the input turbulent spectrum for the modelling of $HCO^+$. As long as the velocity scale at $l_{AD}$, $V(l_{AD})$ is smaller than $V_{abs}$, our result will still hold.

### 7. IMPROVEMENT OVER PREVIOUS STUDIES

As a demonstration, we apply the same technique employed by Paper I to our set of data and discuss its applicability. The measure of velocity dispersion versus size is an important way to characterize the turbulence in molecular clouds. In order to study the hierarchical turbulent structures, Ostriker et al. (2001) coarse-grained their simulated spectral data cubes to different resolutions/beam size and measured the velocity dispersions along all the line of sights as a function of scale as in Figure 4. They showed that the underlying turbulent velocity spectrum can be well-recovered by fitting the lower envelope of such velocity dispersion distributions. This method relies on the idea that the lower envelope of the $\sigma^2$ distribution traces regions that sample the shortest line of sights, such that the change in the beam size dominates the variation in the measured $\sigma^2$. This technique was studied further in detail by Falceta-Gonçalves et al. (2010). They showed that this technique only marginally recovered the actual dispersion in marginally supersonic case (See Figure 2 of Falceta-Gonçalves et al. (2010)). Furthermore, they had showed that in highly supersonic simulations the lower envelope of such velocity dispersion distribution will underestimate the actual velocity dispersion as a function of scale. It can be understood with the fact that supersonic turbulence can easily induce high density contrast. The resultant distribution is therefore biased to denser regions and less susceptible to the velocity distribution. This renders the application of this method to observational data problematic.

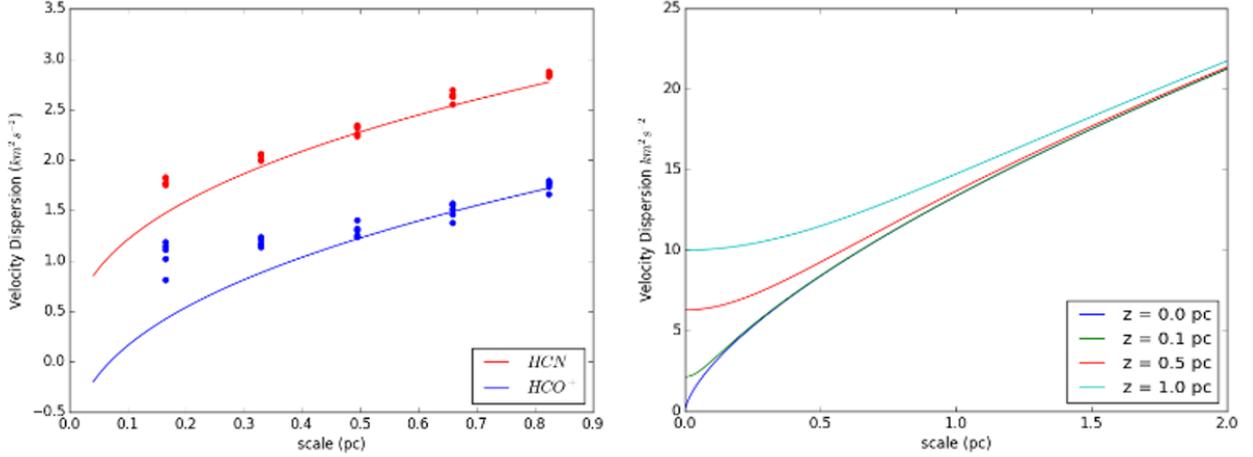

**Figure 4.** Left Panel: The lower envelope of the squared velocity dispersion of $HCN$ and $HCO^+$ are shown in red and blue respectively as a function of Beam size $R$. Our best-fit of the lower envelope is presented in the equation 7. Right Panel: The dependence of $\sigma^2$ as a function of beam size $R$ on the line of sight scale $z$, $\sigma^2 \propto (\sqrt{(2R)^2 + z^2})^m$. As the beam $R$ size decreases, the observed velocity dispersion $\sigma^2$ saturates at $z^m$. This lower envelope technique might not recover the turbulent spectrum well for scale smaller than the line of sight depth.

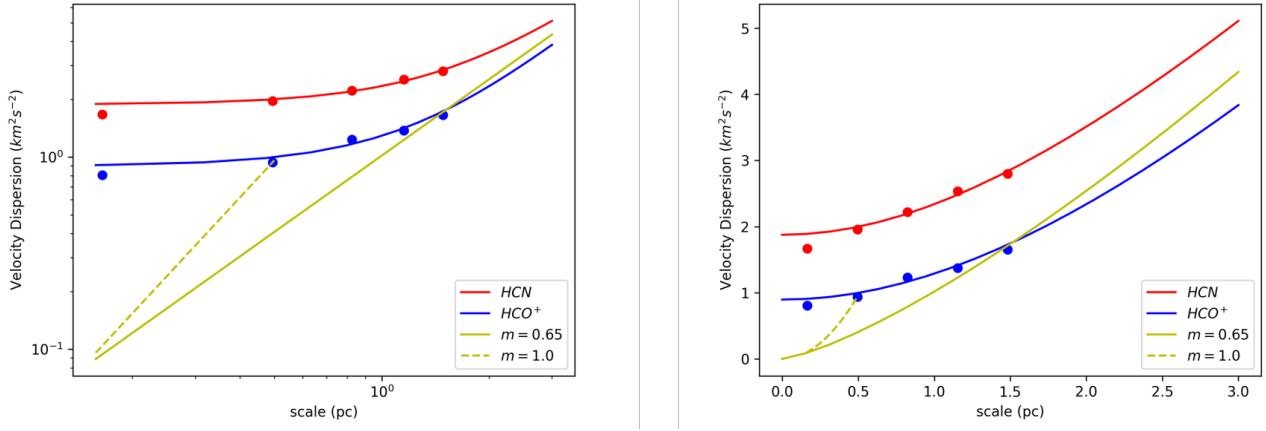

**Figure 5.** The lower envelope fitting accounting for the velocity dispersion contribution from the finite line of sight of the observed cloud in linear scale (Left Panel) and log-log scale (Right Panel) respectively. The filled dots show the lower envelope of the $HCN$ (red) and $HCO^+$ (blue). The solid lines are the respective best-fit from the lower envelope fitting. For reference, we also show here the scaling we derived from VCS, i.e. $\sigma^2 \propto r^{0.655}$ ($HCN$) and $\sigma^2 \propto r^{1.04}$ ($HCO^+$).

We first sample the lower envelope of $\sigma^2$ of $HCN$ as a function of (synthetic) beam size, and locate the pointing where HCN samples the shortest line of sight. We then compare the $\sigma^2$ of $HCO^+$ from these locations. The lower envelope of HCN can be well described by a power law fitting for $R \geq 0.3 pc$ as shown in Figure 4.

$$\begin{aligned}\sigma_{HCN}(R)^2 &= 2.99(\frac{R}{1pc})^{0.393} km^2 s^{-2} \\ \sigma_{HCO^+}(R)^2 &= 2.99(\frac{R}{1pc})^{0.393} - 1.05 km^2 s^{-2}\end{aligned} \quad (7)$$

| | |
|---|---|
| $z_{HCN}$ | $1.12 \pm 0.07 pc$ |
| $z_{HCO^+}$ | $1.48 \pm 0.35 pc$ |
| $C$ | $1.61 \pm 0.09 km^2 s^{-2} pc^{-1.3}$ |
| $D$ | $1.80 \pm 0.8 km^2 s^{-2}$ |

**Table 2.** The list of parameters derived from fitting the lower envelope of velocity dispersions accounting for the finite line of sight depth.

We interpret the deviation of the $\sigma^2$ at $R \leq 0.5pc$ for both $HCN/HCO^+$ as the saturation of the $\sigma^2$ when the beam size falls below the line-of-sight scale. Here we demonstrate the effect of line of sight depth on the measured $\sigma^2$ in Figure 5 with a simple model $\sigma^2 \propto (\sqrt{2R^2 + z^2})^m$, with R being the beam size, z being the line of sight depth and m being the scaling exponent of turbulence. When the line of sight depth $z = 0pc$, the variation of $\sigma^2$ dominates by the change in beam size. This is the ideal scenario where this lower envelope method works best. For models with $z > 0pc$, when the beam size $R \leq z$, $\sigma^2$ only settles on the lower limit set by the line of sight scale z. The scaling of $\sigma^2$ recovers the trend traced by the blue lines for beam size with $R \geq z$.

Figure 5 might also explain why the slopes inferred from this method to observation data ($m \approx 0.36$ (Paper I), $m \approx 0.3 - 0.52$ (Hezareh et al. (2010, 2014))) are consistently shallower than the expected scaling (Kolmogorov: $m = 2/3$; Compressible Turbulence: $m = 1$; This studies with VCS: $m \approx 0.65$)). Considering scales where $R \approx z$, the slope/scaling inferred from this plot is always shallower than the actual spectral index. As observers for most of the time have no accurate information about the line of sight depths, it is hard to determine which regime we are in. Moreover, the inhomogeneous density structures along the line of sight could also affect the lineshape/linewidth. The result of our simplified model assuming homogeneous density distribution indeed resembles the prediction in the sub-sonic simulations (See Figure 2 of (Falceta-Gonçalves et al. (2010))).

Here we also experiment the fitting of the lower envelope accounting for the contributions from the finite line of sight depth with the scaling index, $m = 0.65$ derived from VCS fixed (Figure 5). This allows us to estimate the line of sight depth of these two tracers.

$$\sigma^2_{HCN}(R, z_{HCN}) = C(\sqrt{2R^2 + z^2_{HCN}})^{0.65}$$
$$\sigma^2_{HCO^+}(R, z_{HCO^+}) = C(\sqrt{2R^2 + z^2_{HCO^+}})^{0.65} - D \qquad (8)$$

Free parameters for the fitting and the respective best fits parameters are shown in Table 2. The fitting shows that the line of sight depth estimate $z_{HCN} \approx 1.12 \pm 0.07pc$ is consistent with the cloud scale $L \approx 1.26 \pm 0.11pc$ we derived with VCS. This reinforces our explanation on the saturation of velocity dispersion due to unresolved line of sight depths. Furthermore, we also find that $z_{HCO^+}$ is larger than $z_{HCN}$. This is also consistent with the fact that $HCO^+$ has a lower critical density than $HCN$ (Jansen (1995)), and therefore is capable of exciting emission from a longer line of sight depth. This also explains the larger spatial extent $HCO^+$ has than that of $HCN$ observed in this work (Figure 2, Left Panel) and possibly observations in general star forming regions (Paper I, Hezareh et al. (2014); Houde et al. (2000); Lo et al. (2009)). Since the line-of-sight depth is comparable to the plane-of-the-sky scale, this hints at a flattened, disk-like cloud core geometry. This is a signature of anisotropic collapse induced by dynamically-important B-fields since magnetic force only acts perpendicular to the field lines. The non-spherical nature of star forming cores has been studied extensively by Tritsis et al. (2015) with observation data.

We also examine the sensitivity of the VCS fit of $HCO^+$ to the change in $z_{HCO^+}$. In the previous section, we adopt the best-fit cloud scale of $HCN$ as that of $HCO^+$. This might be oversimplifying as $HCO^+$ is expected to trace a longer line of sight. We repeat the fitting of VCS for $HCO^+$ by replacing $L$ with $z_{HCO^+} \approx 1.48pc$. The optimal fitting gives an $l_{AD} \approx 0.547 \pm 0.213pc$ and is still in agreement with the result we reported $l_{AD} \approx 0.404 \pm 0.132pc$.

## 8. IMPLICATION OF THIS STUDY

### 8.1. *Validity of Chandraskhar-Fermi Method on sub-pc scale*

Chandrasekhar and Fermi (Chandrasekhar & Fermi (1953)) proposed a method to measure B-field strength by assuming the equipartition between turbulent kinetic energy (velocity dispersion) and turbulent B-field energy, which is inferred from thermal dust polarization (hereafter C-F method). However, this assumption might break down for scales $L \leq l_{AD}$, when the B-field decouples from bulk gas flows. Indeed, simulations have shown that the deviation of the estimate with C-F method with the actual B-field strength increases as the magnetic Reynold number decreases (Li et al. (2012)). In specific, the C-F method will over-estimate the actual B-field strength. C-F method therefore should be applied with care depending on the physical conditions of the molecular clouds.

### 8.2. *Possible Resolution to the magnetic braking catastrophe*

The mechanism behind proto-stellar disk formation is still a matter of debate and turbulence has been proposed to be the source of the angular momentum (e.g. Seifried et al. (2012)). B-field directions from cloud cores at 0.1 pc scale and their surrounding inter-cloud medium at 100 pc scale are found to be well aligned (Li et al. (2009)). Recently, this signature of dynamically dominant B-fields has also been observed down to 0.01pc (Li et al. (2015)). Whether the situation extends to the scales of proto-stellar disk formation is unknown, but recent simulation presented by Zhang et al. (submitted) suggested dense cores formed out of an initially sub-Alfvenic parent cloud cannot be highly super-Alfvenic. In other words, turbulent eddies may not be energetic enough to drive disk rotation against magnetic forces. The lack of an effective energy supply to overcome B-fields and form disks is called magnetic braking catastrophe. The observation of $l_{AD}$ at sub-pc scale may shed light on this catastrophe. With the canonical ionization equilibrium $n_i \propto n_n^{1/2}$ (Draine (2011)) and a constant drag coefficient between ions and neutrals $\gamma_d$, the magnetic Reynold number $R_M \propto (n_i n_n l V)/B^2$ should scale roughly to $R_M \propto l^{0.1-0.9}$ which decreases monotonically as one progresses to even smaller scales. If turbulence-driven AD is at work already at sub-parsec scales as presented, a high degree of decoupling between turbulent eddies and the magnetic fields is expected for disk forming kAU scales. This also implies that disks with angular momentum stems from local turbulent eddies within the decoupling scale need not to be aligned with the local B-field (Hull et al. (2014)), which is usually expected in the strong field scenario where flux freezing is assumed.

## 9. CONCLUSION

Building upon the foundations laid in Paper I, we present a new recipe to model turbulent ambipolar diffusion, and apply it to a diffuse quiescent molecular filament in NGC6334. With VCS, we model the turbulent energy spectrum of both ions and neutrals with $HCN$ and $HCO^+(4-3)$ spectral lines. We show that the result is best understood with a break in the ions' turbulent spectrum at $l_{AD} = 0.404 \pm 0.132 pc$. This is consistent with the theory put forward by Paper I that the disturbances in magnetic fields are damped by turbulence-driven AD. We argue that our result is unlikely to be the artifacts from the tracers properties, chemistries or optical depths. Moreover, we show that the method adopted in this work can circumvent the effect of unresolved line of sight depth of the tracers and is better at tracing the actual turbulent spectrum, compared to the technique used in previous works. The discovery of turbulence-driven AD that operates at the sub-pc scale is important, as this challenges the applicability of CF method at the molecular cloud core scale, and might possibly shed light on the magnetic braking catastrophe in the theory of proto-stellar disk formation. More observations would be essential to see if it is a general process in the molecular cloud cores and to constrain its role in the context of star formation.

K.S.T. would like to thank Telemachos Mouschovias for stimulating discussions that helped improve this work. K.S.T. would also like to thank Umemoto Tomofumi for the tutorial on the operation of the ASTE telescope. W.K.L. acknowledges support from the Theoretical Institute for Advanced Research in Astrophysics in the Academia Sinica Institute of Astronomy & Astrophysics. The ASTE telescope is operated by the National Astronomical Observatory of Japan (NAOJ).

This research is supported by four grants from the Research Grants Council of Hong Kong: Early Career Scheme 24300314; General Research Fund 14305717, 14600915, and 14304616.

APPENDIX

A. THE FITTING OF VELOCITY COORDINATE SPECTRUM

The Velocity Coordinate Spectrum (VCS) technique based on the study of the intensity fluctuations/ spectral lines along the velocity axis. The theoretical prediction of such power spectrum based on the assumption of a power law like spectrum for both turbulent velocity and the density distribution. VCS is defined to be the power spectrum of the intensity fluctuations in the velocity coordinate along fixed line of sight:

$$P_{vcs}(k_v) = |\int_{-\infty}^{\infty} I(v) e^{ik_v v} dv|^2, \tag{A1}$$

where $I_v$ is the spectral line intensity at velocity $v$ and $k_v$ is the wavenumber in the velocity coordinate (e.g., $k_v = 2\pi/v$). The expression of VCS is of the following form (Chepurnov et al. (2015))

$$P_{vcs}(k_v) = f^2(k_v) P_0 \int d^3\vec{r} \ C_\epsilon(\vec{r}) g(\vec{r}) d\vec{r} e^{-k_v^2/2D_{vz}(\vec{r})} + N_0 \tag{A2}$$

where f is the Fourier transform of the Maxwellian distribution which depends on temperature, $P_0$ is the spectrum amplitude, $C_\epsilon(\vec{r})$ is the density correlation function, $g(\vec{r})$ is the geometric factor that accounts for the beam pattern and structures along the line of sight, $D_{vz}(\vec{r})$ is the structure function of the turbulent velocity component along the line of sight and $N_0$ is a constant which depends on the noise level.

The temperature dependent term f and the density correlation function term $C_\epsilon$ can be determined independently before the fitting of VCS. The temperature of this filament is of the order $20 - 30K$ (André et al. (2016)). This translates to a thermal velocity of $0.1 km s^{-1}$ for both $HCN$ and $HCO^+$. In Figure 3, it is clear to see that $f^2(k_v, T)$ has a weak functional dependence on $k_v$ in the range we fit our model. The density correlation function can be derived from the study of the 2D spatial power spectrum of the column density map (Lazarian & Pogosyan (2000)). The power spectrum would be in the form of $P_{2D}(k) \propto k^{-3+\gamma}$. The density correlation function is steep when $\gamma < 0$ and shallow when $\gamma > 0$. In fact, that the contribution of the density fluctuations has a weak dependence on the VCS when the density correlation function is steep (Lazarian & Pogosyan (2006)). Here we showed the spatial power spectrum of the column density map from both $HCN$ and $HCO^+$ in Figure 6. These two species showed almost an identical trend which is again an indication of the high correlation between the two species. Since $\gamma < 0$ for both species, the density correlation term $C_\epsilon(\vec{r})$ can be factored out from the integration that follows.

In the optically thin limit, VCS can now take a simpler form

$$P_{vcs}(k_v) = P_0 \int d^3\vec{r} \ g(\vec{r}) e^{-k_v^2/2D_{vz}(\vec{r})} + N_0 \tag{A3}$$

Free parameters needed to model the VCS reduces to $P_0$, $N_0$, $D_{vz}(\vec{r}) = <(v_z(\vec{x}) - v_z(\vec{x}+\vec{r}))^2>$. The structure function $D_v(r)$ and the turbulent energy spectrum can be converted from one to the other through Fourier Transform.

|  | $0 \leq r \leq l_{AD}$ | $r \geq l_{AD}$ |
|---|---|---|
| HCN | $Cr^m(1+\frac{m}{2}(1-\frac{z^2}{r^2}))$ | $Cr^m(1+\frac{m}{2}(1-\frac{z^2}{r^2}))$ |
| $HCO^+$ | $C_{AD} r^{m_{AD}}[1+\frac{m}{2}(1-\frac{z^2}{r^2})]$ | $F + Cr^m[1+\frac{m}{2}(1-\frac{z^2}{r^2})]$ |

**Table 3.** The structure function $D_{vz}(r,z)$ for $HCN$ and $HCO^+$ respectively.

For Komolgorov type turbulence, the spectral index $\kappa$ is given by 5/3. An extra spectral index $\kappa_{AD}$ is introduced for the charged species to account for the dissipation of turbulent energy. However, when the line of sight scales $z$ is no longer comparable with the plane-of-sky scale, the structure functions of the turbulent velocity component $D_{vz}(r,z)$ along the line of sight (Lazarian & Pogosyan (2006)) for ions and neutrals are more complicated and are shown in

$$\begin{aligned} C_{AD} &= m/m_{AD} \quad Cl_{AD}{}^{m-m_{AD}} \\ F &= (m - m_{AD})/m_{AD} \quad Cl_{AD}{}^m \end{aligned} \quad (A4)$$

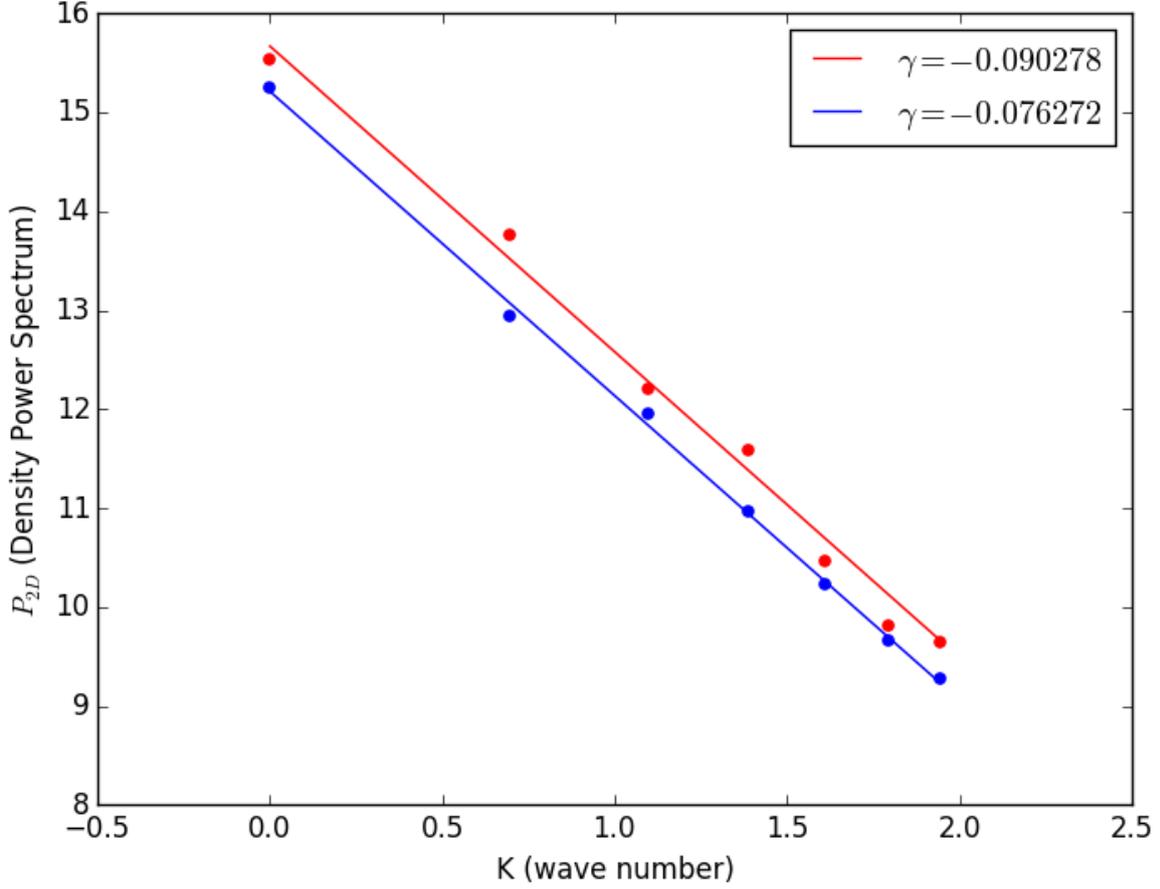

**Figure 6.** The 2D column density power spectrum of $HCN$ and $HCO^+$ in log-log scale. $\gamma < 0$ corresponds to a steep density power spectrum. The density correlation function $C_\epsilon$ can be factored out from the integration of the VCS. This simplifies the fitting procedures.

Table 3. The term $Cr^m$ accounts for the scaling of the velocity dispersion $V$ of the eddies of sizes $r$ while the term $[1 + m/2(1 - z^2/r^2)]$ accounts for the projection of velocity components onto the line of sight direction. The turbulent spectral index $\kappa$ and $\kappa_{AD}$ can be inferred from $m$ and $m_{AD}$ above with $\kappa = m + 1$ and $\kappa_{AD} = m_{AD} + 1$ respectively. $C_{AD}$ and $F$ are set by the boundary condition imposed by the continuity of the turbulent energy spectrum at the scale $l_{AD}$, for a given $C, m, m_{AD}$

The VCS of the observed spectral data cube are first obtained by taking the power spectrum of the spectral data cube along the velocity axis. The average of the VCS from all these different line-of-sight is then taken so to beat down the noise, i.e. $P_{vcs}(k_v) = \frac{1}{N} \sum_{x,y} |\int_{-\infty}^{\infty} I(x, y, v) e^{ik_v v} dv|^2$ (where $N$ is the number of independent line-of-sight) . Before fitting it to the VCS shown above, one also needs to determine whether the density distribution of the observed cloud is in the shallow or steep regime as this affects the VCS fitting form. Parameters required for the fitting of VCS are listed in Table 4. The modelling of VCS begins by first fitting the power spectrum of $HCN$ as it has the least input parameters compared to $HCO^+$ and returns the information of the turbulent spectrum regarding larger scales (e.g.

|  | HCN | HCO$^+$ |
|---|---|---|
| Spectrum Amplitude | **$P_0$** | **$P_1$** |
| Constant (noise) | **$N_0$** | **$N_1$** |
| Spectral Index $l \geq l_{AD}$ | **$m$** | $m$ |
| Velocity Dispersion | **$C$** | $C$ |
| Size of the Cloud | **$L$** | $L$ |
| Spectral Index $l \leq l_{AD}$ |  | **$m_{dis}$** |
| Decoupling Scale |  | **$l_{AD}$** |

**Table 4.** Input parameters for fitting VCS to the power spectrum of $HCN$ and $HCO^+$. The bold face parameters are the free fitting parameters. The fitting of $HCO^+$ is followed by the modeling of $HCN$.

| HCN | | HCO$^+$ | |
|---|---|---|---|
| $C$ | $9.16 \pm 1.47\,km^2 s^{-2}$ | $m_{dis}$ | $1.01 \pm 0.049$ |
| $m$ | $0.655 \pm 0.0935$ | $l_{AD}$ | $0.484 \pm 0.158\,pc$ |
| $L_{cut}$ | $1.26 \pm 0.112\,pc$ | $C_{dis}$ | $7.69 \pm 2.07\,km^2 s^{-2}$ |
|  |  | $D$ | $2.0 \pm 1.35\,km^s s^{-2}$ |

**Table 5.** The best fit result returned from fitting VCS to the power spectrum of $HCN$ and $HCO^+$

$C, m, L$). These parameters are used in the fitting of the spectrum of $HCO^+$. The resultant best fit parameters and their respective one standard deviation error returned from the Scipy Optimize module are listed in Table 5.

## B. WOULD $HCN$ BE BETTER EXPLAINED BY A BROKEN POWER LAW

Ideally, when we fit VCS to our result, we should not pre-assume an underlying form of turbulent spectra for the either species. However, our current implementation of VCS is incapable of fitting a more generalized form of turbulent spectra, i.e. including simultaneously the large scale turbulent spectrum, the break scale and the spectral index beyond the break.

In this section, we will test the robustness of the assumption of our single power-law turbulent spectrum for $HCN$. This is essential as the result constraints the input for the fitting of $HCO^+$. We fit the VCS of $HCN$ for a range of dissipation scale $l_{dis}$ assuming the initial $HCN$ best-fit large scale turbulent spectrum as shown in Table 5. If a single power-law turbulent spectrum is favored, the spectral index beyond the assigned break would be close to the initial large scale power-law spectrum. The result are shown in Figure 7. The red dots are the best fit scaling index for the assumed $l_{dis}$ and are almost identical to our initial large scale spectrum. This verifies our assumption of a single power-law turbulence spectrum for $HCN$. We show alongside the same experiment result with $HCO^+$. The results clearly favors the explanation with a steep turbulent spectrum for $HCO^+$.

## C. THE EFFECT OF NOISE AND RESOLUTION ON VCS

VCS is the power spectrum of the observed spectral lines. We can qualitatively understand the effect of noise on the observed spectrum by adding an additional noise term on top of the true signal. Here we introduce $N(v)$ to mimic the actual observations across the velocity channel. Since $N(v)$ should be uncorrelated with the signal, its cross-integral with the actual signal would therefore vanish. The composite VCS would be left with two terms: the VCS of the clean signal, and the power spectrum of the noise term. Assume the noise term is Gaussian distributed with zero mean and a dispersion of $<|N(v)|^2> = \sigma$. By Parseval's theorem, $<|N(k_v)|^2> = \sigma$. Effectively, the noise term would introduce a baseline on the VCS spectrum. As soon as $P_{vcs}(k_v)$ is comparable to $\sigma$, the VCS spectrum would be buried by the noise term and hence reduce the range available for fitting. Both VCS for ions and neutrals shown in Figure 3 flatten off to constant value at high $k_v$. This reinforces our assumption of a constant noise power spectrum. The effect of noise enters explicitly in the fitting function through the term $N_0$ defined in Table 4.

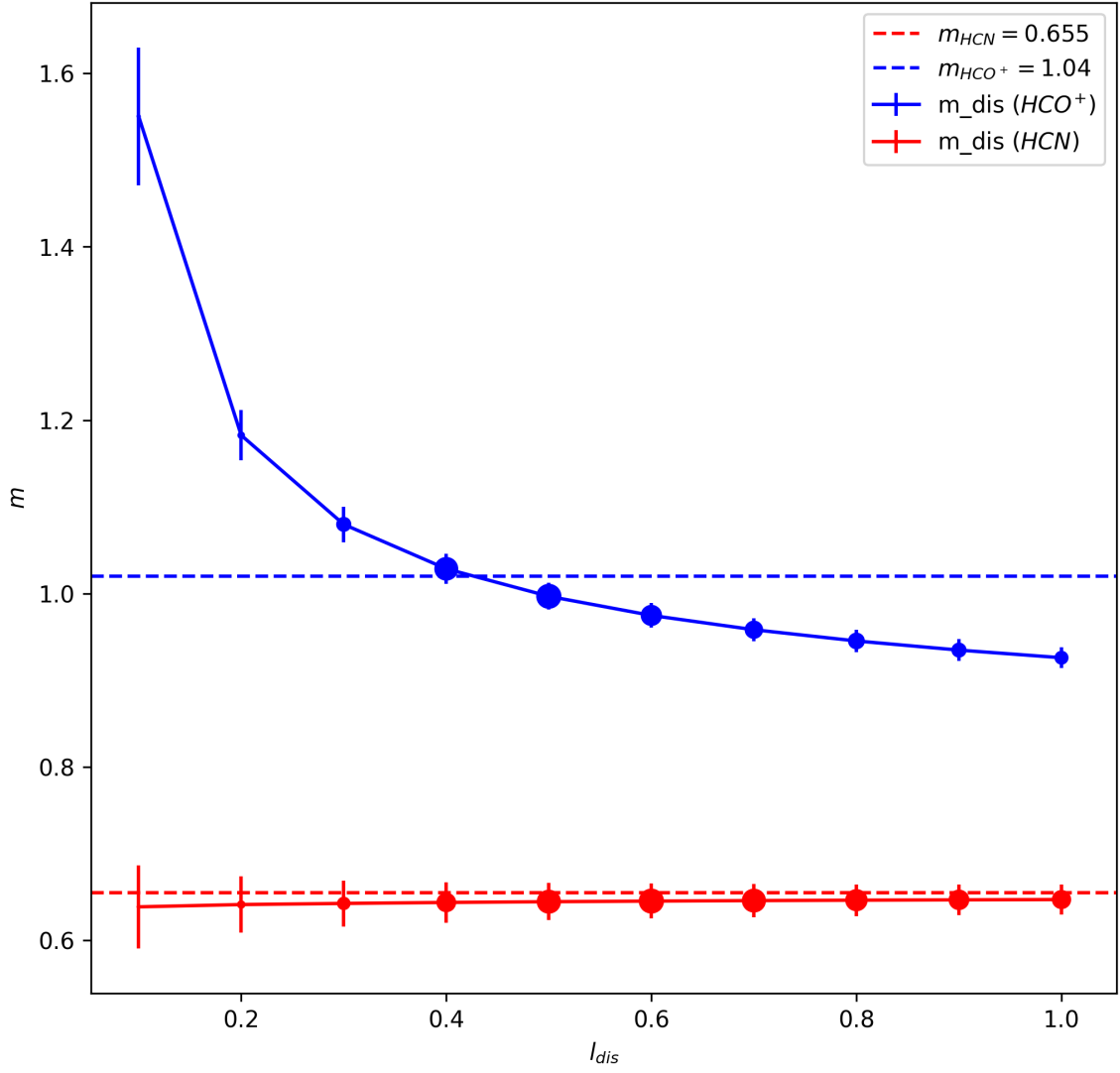

**Figure 7.** If $HCN$ is better fitted by a broken turbulent power law as in $HCO^+$, the best-fit spectral index beyond the break would deviate significantly from the initial best-fit. Results are shown above in red. It is clear that they are close to the initial spectral index. The best fit spectral index for $HCO^+$ is shown alongside for comparison. The diameter of the marker is scaled inversely with the root-mean-squared error between the best-fit VCS and the actual data, i.e. the larger the marker size, the smaller the root-mean-squared error, the better the fitting. This reassures our claims that $HCN$ is better fitted with a power law like turbulent spectrum without a break.

$$\begin{aligned}
P_{vcs,noise}(k_v) &= |\int_{-\infty}^{\infty} (I(v) + N(v))e^{ik_v v} dv|^2 \\
&= \int_{-\infty}^{\infty} I(v_1)I(v_2)e^{ik_v(v_1-v_2)}dv_1 dv_2 + \int_{-\infty}^{\infty} N(v_1)N(v_2)e^{ik_v(v_1-v_2)}dv_1 dv_2 + 2\int_{-\infty}^{\infty} I(v_1)N(v_2)e^{ik_v(v_1-v_2)}dv_1 dv_2 \\
&= P_{vcs}(k_v) + <N(k_v)^2> \\
&= P_{vcs}(k_v) + N_0^2
\end{aligned}$$
(C5)

In the model of VCS, the size of telescope beam is handled explicitly through the geometric factor $g(\vec{r})$ (Chepurnov et al. (2010)). Chepurnov et al. (2010) and Chepurnov et al. (2015) showed that the fitting of VCS on the same data with different resolution give the same fitting results on the turbulent spectra by changing only the geometric factor. This shows the robustness of the techniques on the effect of spatial resolution. In reality, spatial resolution limits the number of independent line-of-sight spectrum, and increases the error bar of the VCS inferred from the data.